\documentstyle[psfig]{mn}
\def\H0{{\it H}$_0$}
\def\Ms{{\it M}$_\odot$}

\def\q0{{\it q}$_0$}

\def\ergps{erg~s$^{-1}$}

\def\Ms{{\it M}$_\odot$}

\def\nH{$N_{\rm H}$} 
\def\psqcm{cm$^{-2}$}

\def\cps{ct\thinspace s$^{-1}$}
\def\Rin{$R_{\rm in}$}
\def\Rout{$R_{\rm out}$}

\def\rg{$r_{\rm g}$}
\def\phpspsqcm{ph\thinspace s$^{-1}$\thinspace cm$^{-2}$}

\title[Iron K emission line in MCG--6-30-15]
{The variable iron K emission line in MCG--6-30-15}
\author[K.~Iwasawa {\it et al.}]
{\parbox[]{6.5in}{K.~Iwasawa$^{1}$, A.C.~Fabian$^1$, C.S.~Reynolds$^1$,
K.~Nandra$^2$, C.~Otani$^3$, H.~Inoue$^4$, K.~Hayashida$^5$, W.N.~Brandt$^1$,
T.~Dotani$^4$, H.~Kunieda$^6$, M.~Matsuoka$^3$ and Y.~Tanaka$^{4,7}$}\\
\\
1: Institute of Astronomy, Madingley Road, Cambridge CB3 0HA\\
2: Laboratory for High Energy Astrophysics, 
Goddard Space Flight Center, Greenbelt, MD 20771, USA\\
3: The Institute of Physical and Chemical Research (RIKEN), Hirosawa, Wako, 
Saitama 351-01, Japan\\
4: Institute of the Space and Astronautical Science, Yoshino-dai, Sagamihara, 
Kanagawa 229, Japan\\
6: Department of Earth and Space Science, Osaka University,
Toyonaka, Osaka 560, Japan\\
6: Department of Astrophysics, Nagoya University, Chikusa-ku, 
Nagoya 464-01, Japan\\
7: Max-Plank Institut f\"ur Extraterrestrische Physik,
Giessenbachstrasse, D-85740 Garching, Germany}

\date{}

\begin{document}

\maketitle

\begin{abstract}
We report on the variability of the iron K emission line in the Seyfert 1
galaxy MCG--6-30-15 during a four-day ASCA observation. The line consists
of a narrow core at an energy of about 6.4 keV, and a broad red wing
extending to below 5 keV, which are interpreted as line emission arising
from the inner parts of an accretion disk. The narrow core correlates
well with the continuum flux whereas the broad wing weakly
anti-correlates. When the source is brightest, the line is dominated by
the narrow core, whilst during a deep minimum, the narrow core is very
weak and a huge red tail appears. However, at other times when the
continuum shows rather rapid changes, the broad wing is more variable
than the narrow core, and shows evidence for correlated changes contrary
to its long time scale behaviour. The peculiar line profile during the
deep minimum spectrum suggests that the line emitting region is very
close to a central spinning (Kerr) black hole where enormous
gravitational effects operate.
\end{abstract}


\section{INTRODUCTION}

ASCA observations have revealed that iron K emission lines in the X-ray
spectra of many Seyfert 1 galaxies are broad (Kunieda 1995; Mushotzky et
al. 1995; Tanaka et al. 1995; Iwasawa et al. 1996; Yaqoob et al. 1995;
Fabian et al. 1995 and references therein). The presence of a fluorescent
iron K$\alpha$ line and a high energy hump above 10 keV in the Ginga
spectra (e.g. Pounds et al. 1990; Nandra \& Pounds 1994) has been
explained by reflection from cold thick material subtending a large solid
angle ($\sim 2\pi$ sr) at the X-ray source. An accretion disk can realize
such a geometry and velocities implied from the resolved linewidths,
which greatly exceed those in the optical broad line region, support it.

A recent long ($\sim 4.5$ d) ASCA observation of the bright Seyfert 1
galaxy MCG--6-30-15 has provided the best-resolved line profile so far
(Tanaka et al. 1995). The iron K$\alpha$ emission line is broad and
skewed to low energies, with little emission above the rest energy of 6.4
keV, close to the line peak, and considerable emission down to 5 keV. The
immediate interpretation of the line is that it originates by
fluorescence in the very inner part of an accretion disk (e.g. within 20
\rg) about a massive black hole, illuminated by a primary X-ray source.
The line shape and skewness are due to the combined effects of Doppler
shifts and gravitational redshift from matter in a deep gravitational
potential well, moving in directions within 30$^{\circ}$ from
perpendicular to the line of sight. Since no other plausible broadening
mechanism appears to work (Fabian et al. 1995), we will discuss the line
properties in the framework of the `diskline' model.

MCG--6-30-15 is a nearby Seyfert 1 galaxy ($z=0.008$), and one of the
best-studied active galaxies in X-ray wavebands (Nandra, Pounds \&
Stewart 1990; Nandra \& Pounds 1994; Fabian et al. 1994; Reynolds et al.
1995). This galaxy usually shows large amplitude X-ray flux changes on
time scales of hours and days. This makes it a good target for studying
the response of the line to continuum changes. However, the line flux is
too low to study short timescales and we can only sensibly comment on
changes, or lack of changes, on timescales of longer than $10^4$~s. The
count rate attributable to the line is $\sim 5\times 10^{-3}$ \cps,
meaning that each detector accumulates about 100 ct per line component
per $4\times 10^4$ s which is a large fraction of an operational day.  In
a simple illumination model we could expect the line to follow the
continuum, with lags for the light crossing time (see e.g., Stella 1990).
Since this time is much shorter than a day (it is $500 r_1 M_7$ s at
$10r_1$ gravitational radii for a black hole of mass $10^7 M_7$\Ms), we
do not expect to detect any lag. As we shall demonstrate, changes are
still observed, the explanation for which is less straightforward.

We detect significant line profile changes during the observation in which the
continuum varied with a maximum amplitude by a factor of 7. The importance of
transverse Doppler and gravitational redshift increase as  the region where the
line originates comes closer to the central hole. If the line arises from an
accretion disk about a black hole, the line shape (Fabian et al. 1989; Laor
1991) and its variations therefore provide information about the  location and
movement of the  line emitting regions. The changes imply that the spatial
distribution of the emission is not constant with time. This is the first step
in mapping the inner accretion flow of an active galactic nucleus.

\section{OBSERVATIONS AND DATA REDUCTION}

MCG--6-30-15 was observed by ASCA (Tanaka et al. 1994)
from 1994 July 23 to 1994 July 27, with the Solid-state Imaging Spectrometer 
(SIS) in
Faint/1CCD mode and the Gas Imaging Spectrometer (GIS) in PH mode.
The most well-calibrated chip was chosen for the observation in each SIS.
Data reduction was performed using the ASCA standard software, FTOOLS and 
XSELECT.


\begin{figure*}
\centerline{\psfig{figure=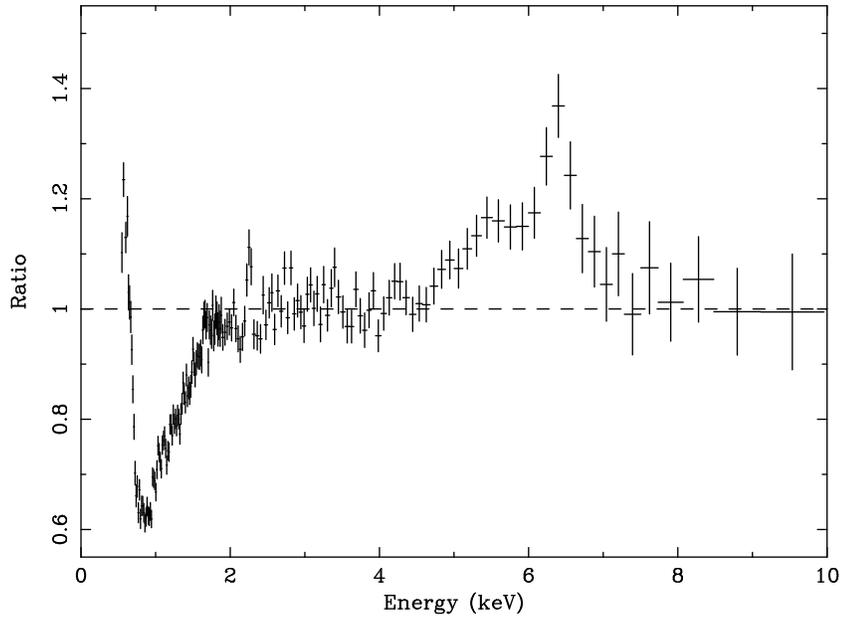,width=0.7\textwidth,angle=270}}
\caption{The ratio of data and model for the averaged 0.4--10 keV spectrum of
MCG--6-30-15. The data are obtained from the S0 detector integrating over the
entire 
long observation (exposure time $\sim 1.7\times 10^5$ s). The model is
a single power-law of a photon index, $\Gamma = 1.96$ modified by cold 
absorption, \nH\thinspace $=6\times 10^{20}$\psqcm,
fitted to the data excluding 
the 0.7--2.5 keV and the 4.5--7.2 keV bands. There is a clear absorption 
feature
around 1 keV, mainly due to OVII and OVIII in the warm absorber, and 
a broad iron K emission line around 6 keV. Note that the effect of the 
warm absorber is restricted to below 2 keV.}
\end{figure*}


\begin{figure*}
\centerline{\psfig{figure=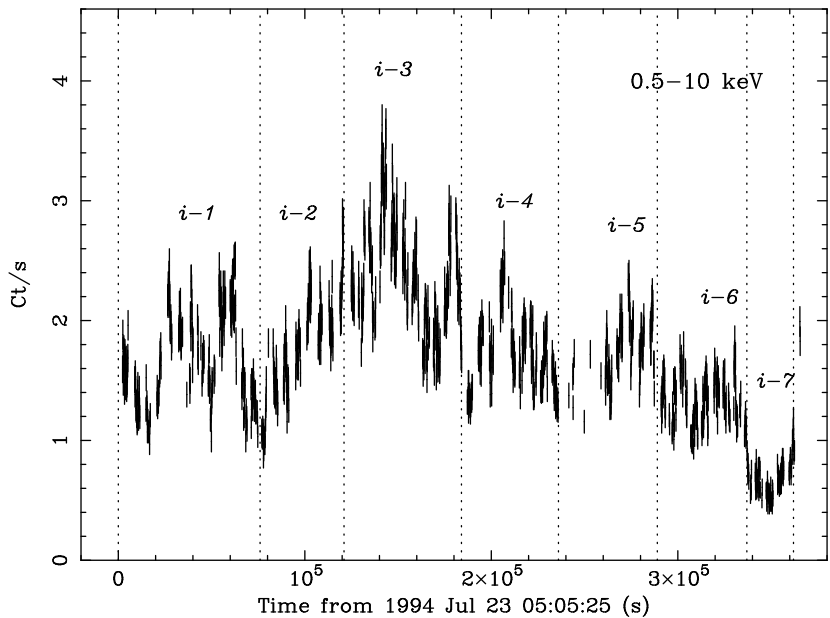,width=0.7\textwidth,angle=270}}
\caption{The 0.5--10 keV light curve from the SIS0. The 
epoch of the start of the light curve is
1994 July 23 05:05:25. Each data bin is averaged over 128 s.
The seven time-intervals used in the section 3.2 are indicated in the figure 
(see Table 1).}
\end{figure*}

The presence of a spectral feature due to partially ionized gas in the
line of sight, the so-called ``warm absorber", established in this source
(Nandra \& Pounds 1992; Fabian et al. 1994; Reynolds et al. 1995), has
been studied in detail for the same observation (Otani et al. 1996). The
spectral distortion of the incident power-law due to this is known to
affect changes below 2 keV (Fig. 1; also e.g., Fabian et al. 1994). To
avoid this complexity, we use the 3--10 keV data for spectral analysis.
A detailed description of the SIS data selection is given by Otani et al.
(1996). The data taken when the source elevation is between 5 and 25
degree above the bright Earth are also used here, although contrary to
standard practice, since our restriction to hard X-ray data renders
these data usable. About 200 ks data are left for spectral analysis, and
the X-ray light curve from a single SIS detector in the 0.5--10 keV band
is given in Fig. 2. For GIS spectra, we applied the standard data
selection criteria; source elevation higher than 5 degree above the Earth
rim, cut-off rigidity larger than 6 GeV c$^{-1}$, and the spacecraft not
in the South Atlantic Anormaly (SAA). The GIS is relatively insensitive
to the line shape, because it has worse spectral resolution than the SIS.
On the other hand, the GIS has better efficiency over the higher energy
band above 5 keV. Due to these complementary characteristics of the two
detectors, we mainly use the SIS data to characterize the line shape, and
the GIS data to help determine continuum slopes. The background data are
taken from a nearby blank field in the same detector field of view for
each dataset with the same time coverage.

\section{RESULTS}
\subsection{Spectral Fitting}

The underlying continuum is fitted by a single power-law absorbed by the
Galactic column density (\nH$\approx 4\times 10^{20}$\psqcm). Even though
a strong high-energy hump in the Ginga spectrum above 10 keV is known in
this source (e.g., Nandra \& Pounds 1994), this reflection component
affects the line flux by only $\sim$5 per cent in observed ASCA spectra.
In the following study, spectra from shorter intervals have fewer counts
than in the total spectrum so the statistical error dominates the
uncertainties on the line intensity.

Modelling of the characteristic shape of the line is basically carried out 
in two ways;
a double gaussian or a diskline model.
In fits with two gaussians, we 
assume line energies and dispersions for the narrow and broad components
to be the same as for the total spectrum (Tanaka et al. 1995),
unless noted otherwise; $E_{\rm B} = 5.5$ keV, $\sigma_{\rm B}=0.64$
keV for the broad component, and $E_{\rm N} = 6.40$ keV, 
$\sigma_{\rm N}=0.15$ keV
for the narrow component, respectively.
The line flux of each component is derived from the SIS spectra, because
the separation of the two components is not appropriate for the GIS energy
resolution.

The diskline model for a Schwarzschild geometry by Fabian et al. (1989)
is used in most cases, assuming a cold accretion disk inclined at 30
degree (Tanaka et al. 1995). Among the parameters of the diskline, the
rest line energy, $E = 6.4$~keV appropriate for K-shell emission from iron
less ionized than FeXVII, and disk inclination, $i = 30$ degree, are
used. The other parameters are the radial emissivity index, $\alpha$,
assuming a power-law type radial emissivity function ($\propto
R^{-\alpha}$) of the line, inner radius of the disk, \Rin, whose
innermost radius of stable orbit for Schwarzshild geometry is 6 \rg, disk
outer radius, \Rout, and normalization of the line. The inner and outer
radii are in unit of gravitational radius, \rg ~= $GM/c^2$. Errors quoted
to best-fit values are at the 90 per cent confidence level for one
interesting parameter while data points in plots have 1$\sigma$ error
bars.

\subsection{Spectra selected in time sequence}

\begin{table}
\begin{center}
\caption{The seven datasets. Count rates in the 3--10 keV band from 
S0 and S1 are given. The time range for each dataset is shown in Fig. 2.}
\begin{tabular}{cccc}
Data & Exposure & 3--10 keV & Time range \\
 & $10^3$ s & \cps & $10^4$ s \\[5pt]
{\it i-1} & 43.4 & 0.274/0.218 & 0.0--7.6 \\
{\it i-2} & 22.2 & 0.273/0.212 & 7.6--12.1 \\
{\it i-3} & 36.1 & 0.390/0.301 & 12.1--18.2 \\
{\it i-4} & 27.2 & 0.296/0.235 & 18.2--23.5 \\
{\it i-5} & 26.1 & 0.295/0.239 & 23.5--28.8 \\
{\it i-6} & 30.4 & 0.243/0.185 & 28.8--33.7 \\
{\it i-7} & 15.2 & 0.152/0.120 & 33.7--36.2 \\
\end{tabular}
\end{center}
\end{table}

We now investigate spectral changes in time sequence. The X-ray source shows
a large flare in the middle of the observation, and a deep minimum of
about 25 ks duration near the end. Including the two extremes, the whole
observing run was divided into seven time intervals with a similar
exposure time ($\sim$ 30 ks) as shown in Table 1 and Fig. 2. The pairs 
{\it i-1} and {\it i-2}, and, {\it i-4} and {\it i-5}, have similar
continuum fluxes to each other.

\begin{figure}
\vspace{-1.5cm}
\centerline{\psfig{figure=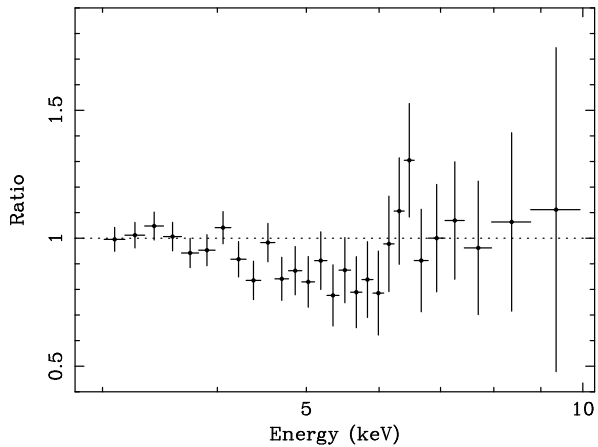,width=0.65\textwidth,angle=270}}
\vspace{-1cm}
\caption{The ratio of `Bright flare'-minus-`Deep minimum' 
data to a power-law with a
photon index of $\Gamma = 2.0$. The power-law is fitted  to the data 
excluding the 
iron-K line band (4--7 keV). This plot demonstrates changes in
spectral shape taking place between
the two extreme flux levels. 
A deficit from 4 keV to 6.4 keV can be seen with a clear peak at
6.4 keV.
This implies an intensity decrease in the broad-wing band and 
an increase in the narrow core band.}
\end{figure}

We present evidence of a spectral change between the `Bright flare' ({\it
i-3}) and the `Deep minimum' ({\it i-7}) in Fig. 3, which is obtained by
taking the ratio of {\it i-7} minus {\it i-3} to a power-law model. The
photon-index of the power-law ($\Gamma = 2.0\pm 0.3$) is obtained by
fitting the data in which the iron line band (4--7 keV) is excluded, and
is consistent with the average continuum slope. If the line flux follows
the continuum and there is no change in line profile or continuum shape
taking place between the two interevals, the subtracted spectrum would
show a line profile similar to the time-averaged line shape as shown in
Fig. 1. A narrow peak at 6.4 keV is therefore understood by an increase
of the narrow core following the continuum change. A clear broad deficit
around 5 keV, however, suggests a decrease of the broad red wing,
contrary to expectation. An alternative explanation for the broad deficit
could be a complicated spectral change in continuum from effects of the
warm absorber and reflection. This possibility will be investigated in
detail for the deep minimum data where also a clear change in the warm
absorber is found (Otani et al. 1996).

\subsubsection{Double-gaussian fits}


\begin{table*}
\begin{center}
\caption{Double gaussian fits to the seven datasets. The S0 and S1 data are
fitted jointly. The best fit photon
indices $\Gamma$ for the SIS spectra
are given with the 90 per cent confidence range 
obtained from the GIS spectra, in parenthesis. 
$\Gamma$ is allowed to vary within the GIS range (see text).
Line energy and dispersion for each gaussian are $E_{\rm N} = 6.40$ keV 
and $\sigma_{\rm N} = 0.15$ keV for the narrow component, 
and $E_{\rm B} = 5.5$ keV and $\sigma_{\rm B}=0.64$ keV for the broad component
through all the fits. 
Note that equivalent widths of the broad component are calculated at the
centroid energy 5.5 keV.}
\begin{tabular}{cccccccc}
Data & $\Gamma$ & & $I_{\rm N}$ & $EW_{\rm N}$ & $I_{\rm B}$ & $EW_{\rm B}$ &
$\chi^2$/dof \\
&&&$10^{-5}$\phpspsqcm & eV & $10^{-4}$\phpspsqcm & eV & \\[5pt]
{\it i-1} & 2.00 & (1.96--2.05) & $4.00^{+1.59}_{-1.63}$ & $114^{+45}_{-46}$ &
$0.88^{+0.31}_{-0.31}$ & $189^{+68}_{-67}$ & 398.8/471 \\
{\it i-2} & 1.92 & (1.82--1.99) & $3.95^{+1.82}_{-1.89}$ & $110^{+52}_{-54}$ &
$0.95^{+0.36}_{-0.49}$ & $200^{+78}_{-106}$ & 381.7/415 \\
{\it i-3} & 2.02 & (1.91--2.02) & $7.50^{+2.07}_{-2.10}$ & $149^{+42}_{-43}$ &
$0.51^{+0.40}_{-0.48}$ & $76^{+61}_{-72}$ & 489.6/499 \\
{\it i-4} & 2.01 & (1.94--2.07) & $5.36^{+2.28}_{-2.38}$ & $143^{+61}_{-64}$ &
$1.02^{+0.58}_{-0.61}$ & $203^{+117}_{-123}$ & 349.3/415 \\
{\it i-5} & 1.97 & (1.97--2.08) & $4.91^{+2.74}_{-1.90}$ & $128^{+75}_{-52}$ &
$0.99^{+0.44}_{-0.42}$ & $193^{+87}_{-83}$ & 324.9/409 \\
{\it i-6} & 1.86 & (1.81--1.94) & $3.58^{+2.05}_{-1.94}$ & $109^{+64}_{-61}$ &
$0.99^{+0.51}_{-0.49}$ & $240^{+123}_{-118}$ & 357.0/394 \\
{\it i-7} & 1.75 & (1.67--1.96) & $1.64^{+2.24}_{-1.64}$ & $81^{+112}_{-81}$ &
$1.56^{+0.40}_{-0.56}$ & $605^{+145}_{-203}$ & 158.1/171 \\[5pt]
\end{tabular}
\end{center}
\end{table*}


\begin{figure}
\centerline{\psfig{figure=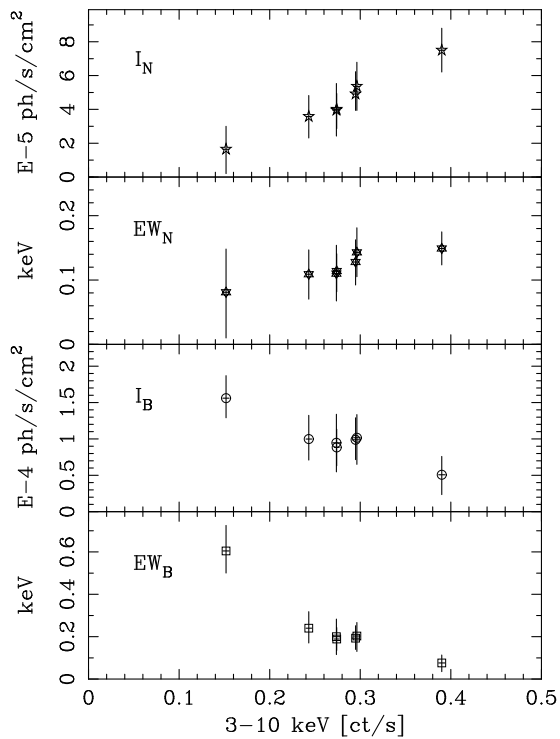,width=0.85\textwidth,angle=270}}
\caption{Plot of intensities and equivalent widths of
the narrow and broad
components as a function of the 3--10 keV count rate (from S0).
The narrow component correlates positively with the continuum flux whereas
the broad component is a decreasing function of continuum.
The correlation coefficient is 0.983 for the narrow component intensity
 ($I_{\rm N}$)
and 0.960 for the broad one
($I_{\rm B}$), implying statistical 
significances above 99.9 per cent for both.
A constant equivalent width for the narrow component cannot be ruled out 
on statistical grounds.}
\end{figure}


\begin{figure}
\centerline{\psfig{figure=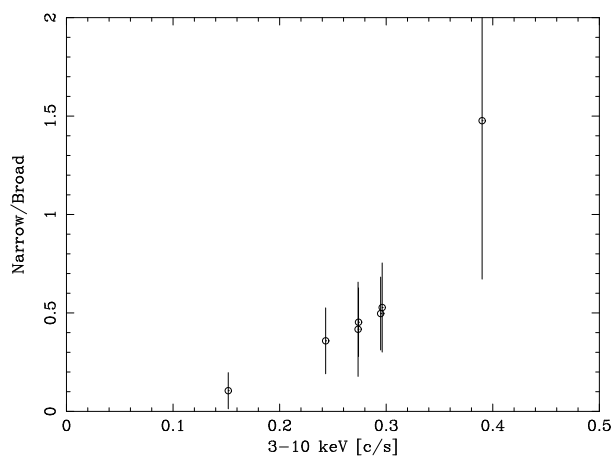,width=0.5\textwidth,angle=270}}
\caption{Ratio of the broad and narrow components as a function of the 
3--10 keV count rate. Intensities of both components are from 
the double gaussian fits in Fig. 4.}
\end{figure}


\begin{figure}
\centerline{\psfig{figure=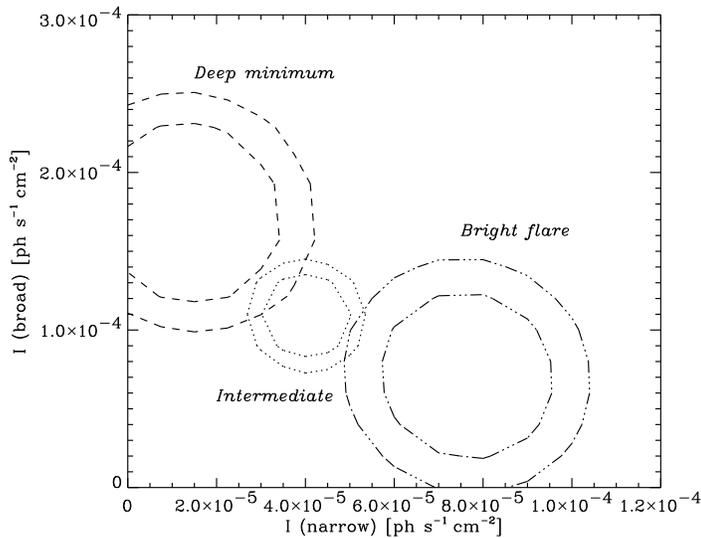,width=0.5\textwidth,angle=0}}
\vspace{0.8cm}
\caption{Contours of the broad versus narrow component intensities for
the Bright flare ({\it i-3}; dash-dot line), Intermediate 
({\it i-1, i-2, i-4, {\rm and} i-5}; dotted line), and Deep minimum 
({\it i-7}; dashed line). Contour levels are at
68 and 90 per cent. Line profiles of the three data sets are shown in Fig. 7.}
\end{figure}


\begin{figure}
\centerline{\psfig{figure=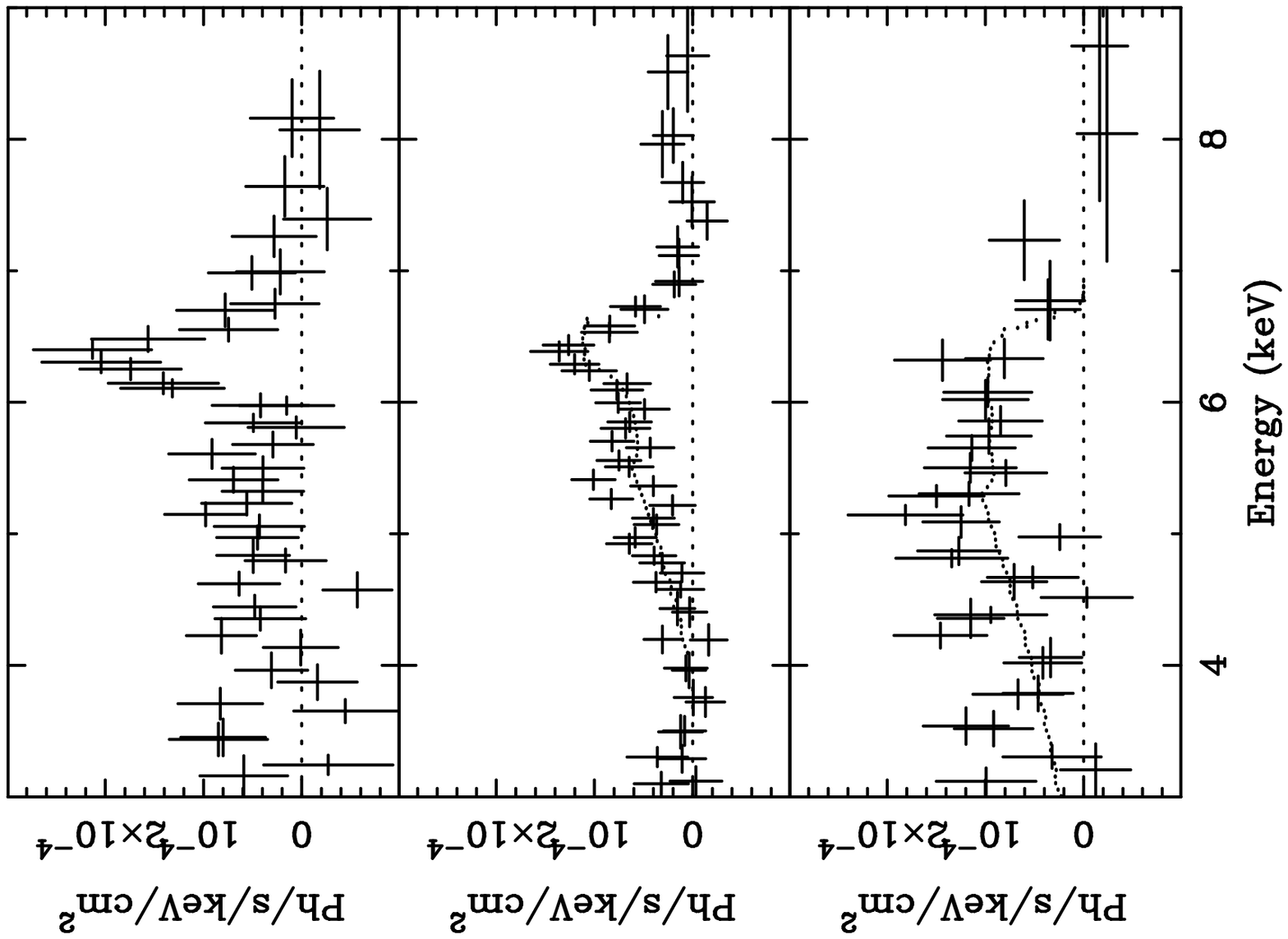,width=0.8\textwidth,angle=270}}
\caption{Line profiles corrected for detector response for interval 3
(the bright flare; {\it Upper panel}), {\it i-1}, {\it i-2}, 
{\it i-4}, and {\it i-5} summed
(intermediate flux intervals; {\it Middle panel}), and interval 7 
(the deep minimum; {\it Lower panel}).  
The line-shape changes between the three flux levels are clearly seen.
The best-fit diskline models for the intermediate flux data and the deep 
minimum data are indicated with dotted lines. The model for the deep minimum 
is L91 with line energy of 6.4 keV as listed in Table 3.}
\end{figure}

The results of the double gaussian fit to the seven datasets are given in
Table 2. Due to the greater efficiency above 5 keV of the GIS, the
continuum slopes are best determined by the GIS spectra. Therefore, in
fitting the SIS data, the photon index is constrained to range within the
90 per cent confidence limits obtained from the GIS result. The photon
indices are very similar from {\it i-1} to {\it i-5} at around
$\Gamma\approx 1.98$. A flatter continuum is suggested for {\it i-6} and
{\it i-7} where the continuum flux is low. A slope change there is
however statistically insignificant. The line energy of the narrow
component fluctuates slightly around the rest energy of 6.4 keV from
dataset to dataset, if allowed to be a free parameter (e.g., $6.55\pm
0.15$ keV in {\it i-5}), but it is always consistent with 6.4 keV; the
precise value does not significantly affect the line intensities
obtained.

Fig. 4 shows the behaviour of the broad and narrow components as a
function of the continuum flux. The intensities of the two components
show the opposite behaviour. The narrow component is correlated well with
the continuum flux. On the other hand, the broad component shows a weak
anti-correlation. As a result, the narrow core is more variable than the
broad wing, though varying in the opposite manner.

The ratios of the broad and narrow components in the bright flare and the
deep minimum data are clearly different from those of the other five data
sets at intermediate flux levels (Fig. 5). This indicates that the line
profile has changed significantly at least in the two time intervals. The
equivalent width of the narrow component is statistically constant
through all the data sets. If the narrow component maintains a constant
equivalent width (EW), then the changes of line profile are due to the
broad red wing since the EW of the broad component in {\it i-3} and {\it
i-7} are significantly different from the rest.

To clarify the changes in intensity of the narrow and broad components
and the ratio of the two, we also show a contour plot of the broad versus
narrow components for the bright flare ({\it i-3}), the deep minimum
({\it i-7}), and a summed dataset of {\it i-1}, {\it i-2}, {\it i-4}, and
{\it i-5}, obtained from the double gaussian fits (Fig. 6). The four
datasets for the summed intermediate data have similar continuum flux
levels and spectral shape as seen in Fig. 4 and Fig. 5; the {\it i-6}
data set is dropped here because of its slightly flatter spectral slope.

We present the line profiles obtained from the bright flare, the deep
minimum, and the summed dataset from the intermediate flux level (Fig.
7). The three line profiles have been fitted by the diskline model,
details of which are given below.

\subsubsection{Intermediate flux data ({\it i-1}, {\it i-2}, 
{\it i-4}, and {\it i-5})}

When the continuum flux is at an intermediate level, the line shape shows
both a narrow core and a broad red wing, similar to the time-averaged
line in Tanaka et al (1995). The double-gaussian fit to the line profile
integrated over {\it i-1}, {\it i-2}, {\it i-4}, and {\it i-5} gives the
contour of narrow versus broad component intensities in Fig. 6 and
the narrow/broad component ratio of $0.33\pm 0.10$. The line is well
fitted by the diskline model with $\alpha= 3.0\pm 1.0$,
\Rin $= 7.6\pm 1.4$ \rg, \Rout = $31^{+30}_{-8}$ \rg, and a line flux of
$1.35^{+0.28}_{-0.23}\times 10^{-4}$ \phpspsqcm ~($\chi^2 = 597.4$ for
642 degrees of freedom). The equivalent width of the line is $EW =
402^{+83}_{-68}$ eV. A combination of relatively large \Rout ~and steep
$\alpha$ obtained here could describe the averaged profile over the four
intervals.

\subsubsection{The bright flare phase (i-3)}

The line shape in this spectrum raises difficulties in 
the diskline fit, due to a strong narrow core.
As the double gaussian fit shows, the broad red wing is relatively fainter than
in other intervals and its EW of 76~eV is significantly smaller than 
the others.

The narrow line shape can be fitted by a diskline model with a very large 
\Rout ~(e.g. $\sim 1000$ \rg ~when $\alpha = 1$), which is indistinguishable
from a single gaussian in the present data. The quality of the fit is
worse than the double gaussian model by $\Delta\chi^2\approx 4$. However,
problems in this model are not only that the red wing is no longer
present in the profile and that the big jump in \Rout ~by a factor of 50
or more is unlikely, but that the continuum becomes only a few times
brighter. Even if we take an extremely flat or even negative emissivity
index (e.g., $\alpha = -3$), it does not work.

We therefore introduce a phenomenological model of the diskline plus an
additional gaussian to fit the line profile. This model, in which the
diskline has the parameters of the `intermediate flux' dataset, improves
the quality of fit. The fraction of the extra narrow component modelled
by a gaussian is now about one half the total line flux.

Since the EW of the narrow component implied from the double-gaussian fit
is not significantly different from that in the other intervals (Table 2
and Fig. 4), the peculiar line profile characterized by a large
narrow-core/red-wing ratio (Fig. 5) is probably due to the suppression of
the red wing. This may imply that the inner radii are less important for
line production or the line emission from the red (receding) side of the
disk is missing.


\begin{table*}
\begin{center}
\caption{Diskline fit to the {\it i-7} spectum. `F89' is for a Schwarzschild
geometry calculated by Fabian et al. (1989) whilst `L91' is for a Kerr geometry
by Laor (1991). \Rin ~and \Rout are set to the minimum value for each geometry
(6\rg ~for Schwarzschild geometry, and 1.25\rg ~for Kerr geometry),
and 15.5\rg, respectively.
The disk inclination is assumed at 30 degree.
$\Gamma$ is a 
free parameter within the range given in Table 2.}
\begin{tabular}{ccccccccc}
Model & $E$ & $\alpha$ & \Rin & \Rout & $I_{\rm line}$ & $EW$ & 
$\Gamma$ & $\chi^2$/dof \\
& keV & & \rg & \rg & $10^{-4}$\phpspsqcm & eV & & \\[5pt]
F89 & 6.4 & $3.2\pm 3$ & 6 & 15.5 & $1.69^{+0.51}_{-0.49}$ & 
$796^{+240}_{-231}$ & 1.75 & 164.2/171 \\
L91 & 6.4 & $2.7^{+0.8}_{-1.2}$ & 1.24 & 15.5 & $2.57^{+1.03}_{-0.87}$ &
$1260^{+505}_{-427}$ & 1.75 & 159.5/171 \\
L91 & 6.7 & $3.4^{+0.6}_{-0.8}$ & 1.24 & 15.5 & $3.44^{+1.21}_{-1.07}$ & 
$1370^{+480}_{-430}$ & 1.78 & 158.0/171 \\
\end{tabular}
\end{center}
\end{table*}


\begin{table*}
\begin{center}
\caption{Fits to the 0.4--10 keV spectrum of {\it i-7} including the
warm absorber and reflection.
Models 1 and 2 do not have any iron line component. 
The warm absorber is computed using CLOUDY (Ferland 1991); parameters are
the 
equivalent hydrogen column density ($N_{\rm W}$)
and ionization parameter ($\xi = L/nR^2$) of the ionized material.
The reflection spectrum from a cold slab inclined by 30 degree is 
computed following Lightman \& White (1988), allowing $\Omega/2\pi$ to
vary. The diskline model of Laor (1991) for Kerr geometry is
fitted here using the  
rest line energy of 6.4 keV and the same inner and outer radii 
of the disk as those in Table 3.
The radial line emissvity index ($\alpha$) and normalization ($I$) are free 
parameters.}
\begin{tabular}{cccccccc}
Model & & \multicolumn{2}{c}{Warm absorber} & Reflection & 
\multicolumn{2}{c}{Fe K line} & $\chi^2$/dof \\[5pt]
& $\Gamma$ & log$N_{\rm W}$ & $\xi$ & $\Omega/2\pi$ & & & \\
& & cm$^{-2}$ & erg\thinspace cm\thinspace s$^{-1}$ & & & & \\[5pt]
1 & $1.79^{+0.02}_{-0.03}$ & 22.43 & 78.96 & --- & \multicolumn{2}{c}{---} 
& 534.6/445 \\[5pt]
2 & $1.91^{+0.03}_{-0.04}$ & 22.16 & 51.05 & $12^{+4}_{-4}$ & 
\multicolumn{2}{c}{---} & 509.2/444 \\[5pt]
& & & & & \multicolumn{2}{c}{Double-gaussian} & \\
& & & & & $I_{\rm N}$ & $I_{\rm B}$ & \\
& & & & & \phpspsqcm & \phpspsqcm & \\
3 & $1.89^{+0.03}_{-0.05}$ & 22.18 & 52.72 & $2.4^{+4.0}_{-2.4}$ &
$2.50^{+2.88}_{-2.42}\times 10^{-5}$ & $1.83^{+0.66}_{-0.49}\times 10^{-4}$ &
484.9/442 \\[5pt]
& & & & & \multicolumn{2}{c}{Diskline\thinspace (L91)} & \\
& & & & & $\alpha$ & $I$ & \\
& & & & & & \phpspsqcm & \\
4 & $1.92^{+0.05}_{-0.04}$ & 22.14 & 49.04 & $3.7^{+4.2}_{-3.3}$ & 
$2.6^{+0.9}_{-1.3}$ & $2.72^{+1.14}_{-0.92}\times 10^{-4}$ & 482.3/442 \\[5pt]
\end{tabular}
\end{center}
\end{table*}


\begin{figure*}
\centerline{\psfig{figure=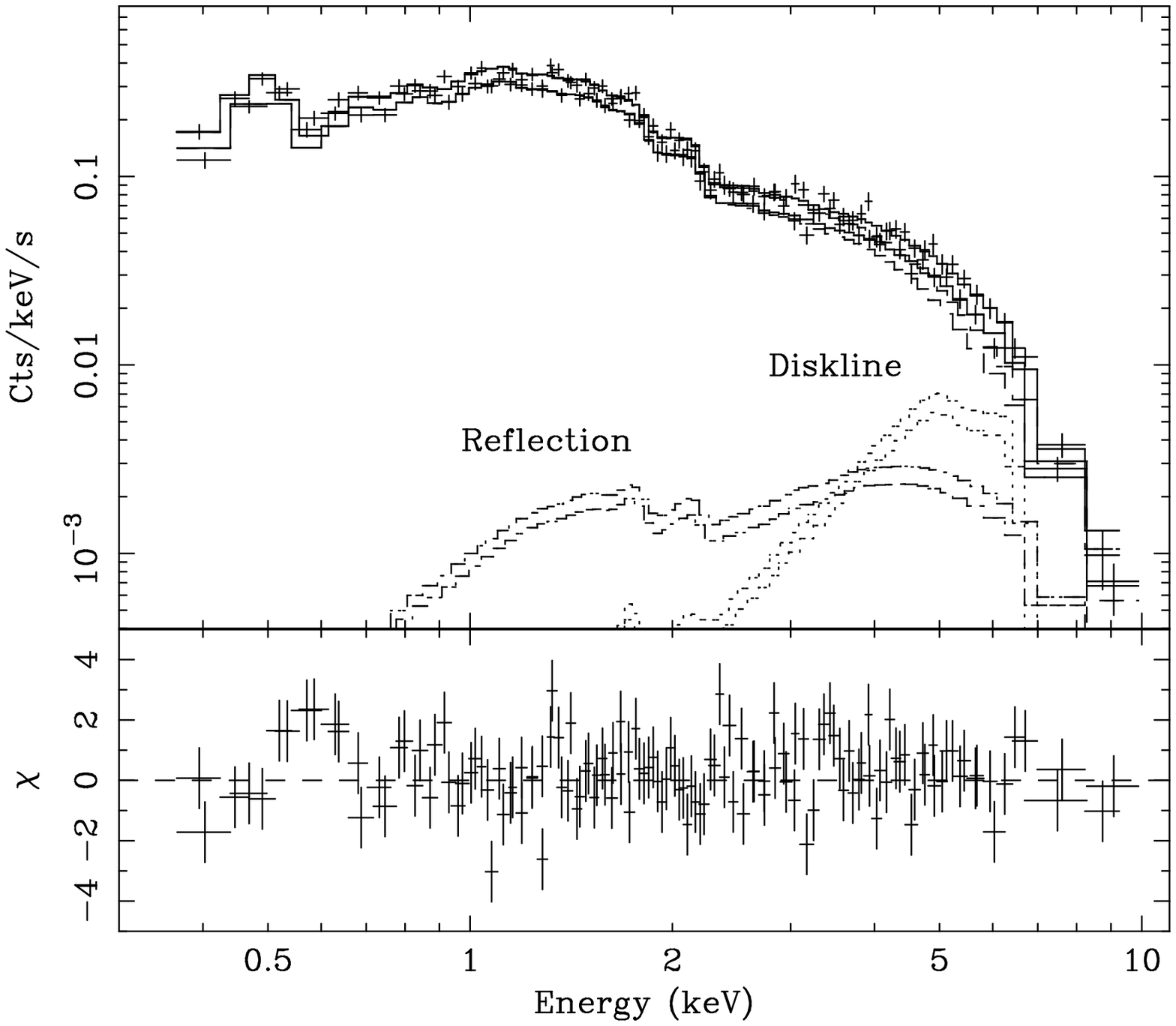,width=0.85\textwidth,angle=270}}
\vspace{-2cm}
\centerline{\psfig{figure=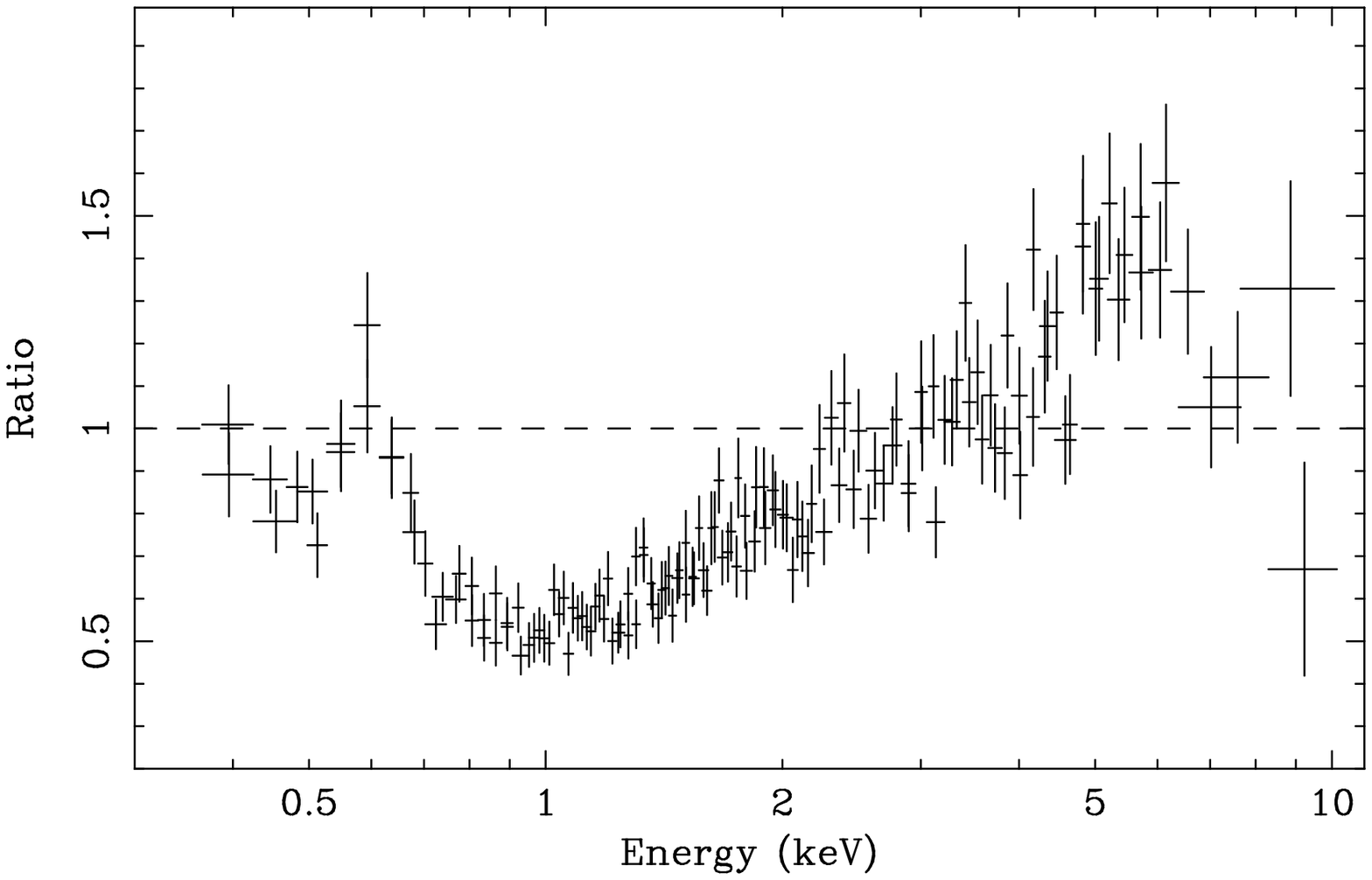,width=0.85\textwidth,angle=270}}
\caption{{\it Upper panel:} 
The 0.4--10 keV SIS spectrum of MCG--6-30-15 during the deep
minimum ({\it i-7}). The data are fitted by a power-law modified by the
warm absorber computed by CLOUDY (Ferland 1991), a reflection spectrum
from cold
material (Lightman \& White 1988), and a diskline model for a Kerr metric
from Laor (1991). The best fit parameters are shown in Table 4 and the model 
spectrum folded through the X-ray telescope and the detector response of ASCA
is displayed in the figure. The residuals at 0.6 keV are most likely an 
instrumental artefact. {\it Lower panel:} 
Ratio of the data and continuum (power-law plus reflection). The plot is 
produced by setting the normalizations of the line and column density
of the warm absorber to zero after making the best fit with the above model.}
\end{figure*}

\subsubsection{The deep minimum phase (i-7)}

In contrast to the flare spectrum, the {\it i-7} spectrum shows the
opposite appearance. Modelling with a double gaussian shows no
significant detection of the narrow line component around 6.4 keV (see
Table 2 and Fig. 4), but within the uncertainities its EW can remain
constant at about 120 eV. The hypothesis of no emission line is, however,
unlikely. A single power-law fit leaves a large negative discrepancy in
the data above 7 keV, relative to the model spectrum. If we introduce a
sharp or a smeared edge, such as is seen in the spectra of Galactic black
hole candidates (Ebisawa 1991), the fit requires an unreasonably flat
spectral slope ($\Gamma\sim 1.2$), and an extremely deep edge ($\tau\sim
1$) at 7 keV. Extrapolation of such a continuum to the low energy band
below 3 keV is also incompatible with the data. This model is therefore
ruled out, and we conclude that there is a strong broad component.

Since a huge red wing may be primarily produced in the innermost parts
of the disk, we set the minimum value for the inner radius as \Rin = 6
\rg, the last stable orbit in the Schwarzschild geometry. The best-fit
emissivity index is then $\alpha = 3.2\pm 3$, but the quality of the fit
is significantly worse than the double gaussian fit ($\Delta\chi^2\approx
-6$), mainly due to the fact that the model cannot explain the large red
wing. A solution for this problem is obtained by introducing the Kerr
geometry of a spinning black hole, which allows a smaller radius for the
last stable orbit (1.24 \rg), following the line profiles calculated by
Laor (1991). This model provides an improved fit with a very similar
$\chi^2$ value to that for the double gaussian fit (Table 3). In this
fit, the radial emissivity index is considerably well constrained
($\alpha = 2.7^{+0.7}_{-1.2}$). The line intensity is found to be even
larger than the total value for the double gaussian fit by $\sim 40$ per
cent.

As noted before (for Fig. 3), a strong red wing could however be due to
complicated continuum effects from the warm absorber and reflection
rather than the change in the line, since we have modelled the continuum
by a simple power-law modified by only Galactic absorption while the warm
absorber increased over this interval (Otani et al. 1996). We examine
such a possibility here by fitting the spectrum of the whole energy range
(0.4--10 keV) with a model including the warm absorber and reflection
continuum. To avoid possible contamination of the low energy data by
reflection from the sunlit Earth, data taken when the source is at an
elevation angle below 25 degree from the bright Earth rim are discarded.
The absorption column density of cold matter is fixed at the Galactic
value (\nH $= 4\times 10^{20}$\psqcm).

The warm absorber in MCG--6-30-15 is usually described with two edges due
to OVII at 0.73 keV and OVIII at 0.92 keV (Fabian et al. 1994; Reynolds
et al. 1995; Otani et al. 1996). Even though the OVIII absorber is
pronounced in this interval, the edge model for OVIII alone does not
reach up to 3 keV, and absorption by the other elements is needed if the
continuum above 3 keV is to be modified. We therefore model a
multi-element warm absorber using a photoionization model computed with
the CLOUDY code (Ferland 1991). This gives a good fit to the present data
with a column density $N_{\rm W}\approx 1.6\times 10^{22}$
\psqcm ~and ionization parameter $\xi\approx $49--53.
The warm absorber with these parameters reduces the flux of a power-law
continuum of $\Gamma = 1.9$ by about 10 per cent at 3 keV. A power-law
modified by the warm absorber alone does not describe the data well
because it leaves a broad bump around 5 keV (Model 1 in Table 4).

Next we investigate the effect of adding an additional reflection
component. The Compton reflection spectrum from cold matter computed in
the form of Lightman \& White (1988) is used here. We continue to assume
that the inclination of the reflecting matter is 30 degree, and keep the
solid angle of the subtending reflection matter at the X-ray source, in
units of $2\pi$~sr (i.e. solid angle of the reflecting material
$\Omega/2\pi$) as a free parameter. As mentioned before, the reflection
spectrum where $\Omega/2\pi = 1$, expected from normal reflection from
an accretion disk, has little affect on the broad line. To assess whether
pronounced reflection can explain the broad feature, this parameter is
allowed to be above the physically reasonable value, i.e., $\Omega/2\pi >2$.

We first fit the 0.4--10 keV spectrum of {\it i-7} by this warm absorber
plus reflection model with no iron K emission line (Model 2 in Table 4).
Fitting the S0 and S1 detectors jointly gives $\Gamma =
1.92^{+0.2}_{-0.5}$ and $\Omega/2\pi = 12\pm 4$ with $\chi^2 = 509.2$ for
444 degrees of freedom. Despite the fit giving an extremely large value
for $\Omega/2\pi$, there is still a broad line-like residual. This is
because the broad feature is too sharp to be explained by the reflection
spectrum. Also data above 7 keV are far below the model, implying that
the predicted reflection hump is too strong. We therefore conclude
that the warm absorber and pronounced reflection cannot explain the broad
line feature totally, even if unusually strong reflection is considered.
In fact, significantly better fits are obtained adding a double-gaussian
model or a diskline model for a Kerr black hole (Model 3 and 4 in Table
4). Results of these fits are summarized in Table 4.

The best-fit line intensities are consistent with those obtained from the
previous fits to the 3--10 keV data. In the fits including the diskline
model for the broad iron line, constraints on the strength of reflection
are poor and no strong reflection is necessary for a good fit to the
data. The equivalent width of the line from the diskline (L91 in Table 4)
against the power-law plus reflection continuum is still large ($EW =
1.28^{+0.53}_{-0.43}$ keV). The data of {\it i-7} in the 0.4--10 keV band
and the best-fit model including the warm absorber, reflection, and the
diskline are shown in Fig. 8.

In summary, the warm absorber and the reflection slightly modify the
3--10 keV continuum, but do not affect the broad line result seriously.
Thus the strong broad red wing of the line in {\it i-7} spectrum appears
to be real from the above inspection, and the broad deficit seen in Fig
3. is not due to continuum effects but to a change in the broad line. The
line profile dominated by a broad red wing may be produced in regions
very close to a central hole where the line is modified seriously by the
gravitational redshift. The unusually large equivalent width may then be
a problem. However, enhancement of reflection due to the returning
radiation of the disk (Cunningham 1976) may be relevant here (see section
4).

\subsection{Intensity-sorted spectra}

We have found significant changes both in intensity and line profile when
the data are sorted in time sequence. They are evident in the bright flare
and the deep minimum datasets but not in the other intervals, mainly
because of similar averaged continuum levels. That study follows the
behaviour of the line averaged over each selected time interval of a few
$10^4$ s. However, the continuum changes on shorter time scales ($< 10^4$
s). If the line follows the continuum with a very short time lag,
intensity-sorted spectra should show the response on short time scale of
the line.

We first investigate spectra obtained by intensity-sorting the whole
dataset (section 3.3.1). However, dramatic changes in the line associated
with large continuum variations on longer time scales, such as for {\it
i-3} and {\it i-7}, confuse the result. Excluding the bright flare and
the deep minimum, changes in the X-ray source are in the intermediate
range ($\sim$1--3 \cps in the 0.5--10 keV band in Fig. 2) and on short
time scales less than $10^4$ s. To clarify the response of the line as
close as possible to faster continuum changes in the intermediate flux
range, we now examine the intensity-sorted spectra excluding the two
time intervals showing the longer time-scale changes (section 3.3.2). All
spectral fits are performed using the double gaussian model.

\subsubsection{The whole dataset}

\begin{table}
\begin{center}
\caption{Intensity-sorted data sets of {\sl H}$_1$ (high), {\sl M}$_1$ (medium)
and {\sl L}$_1$ (low) intensity ranges for whole SIS data.
Data are sorted according to the count rate in the 0.5--10 keV
S0 light curve. Selected ranges of count rate are given in the second column.
Averaged count rates from each S0 and S1 detector in the 3--10 keV band
over the integration time are indicated.}
\begin{tabular}{cccc}
Data & Range & 3--10 keV & Exposure \\
     & \cps & \cps & $10^3$ s \\[5pt]
{\sl H}$_1$ & $> 2.2$ & 0.413/0.318 & 45.2 \\
{\sl M}$_1$ & 1.6--2.2  & 0.320/0.253 & 72.6\\
{\sl L}$_1$ & $< 1.6$ & 0.223/0.178 & 74.2 \\
\end{tabular}
\end{center}
\end{table}

\begin{figure}
\centerline{\psfig{figure=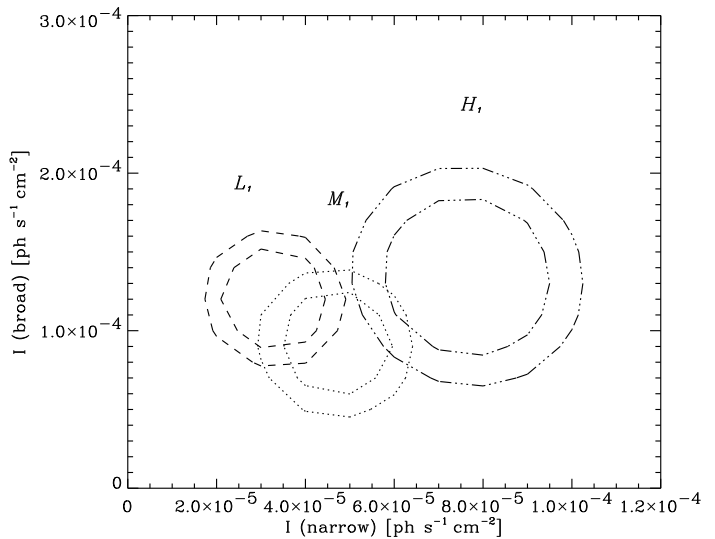,width=0.5\textwidth,angle=0}}
\vspace{0.8cm}
\caption{Contours of intensities between the broad and narrow line components 
for the intensity-decomposed data from whole data. The contours are plotted
in dash-dot line for the {\sl H}$_1$, dotted line for the {\sl M}$_1$, 
and dashed line for the {\sl L}$_1$ data.
The contour levels are at 68 and 90 per
cent for the two interesting parameters.} 
\end{figure}

We decompose the whole data into three flux ranges as shown in Table 5;
{\sl L}$_1$ (low), {\sl M}$_1$ (medium), and {\sl H}$_1$ (high), 
using the count rate
in the 0.5--10 keV in one time bin is averaged over each 128 s. This
decomposition was made using the S0 light curve (Fig. 2), and data from
S1 are extracted from the same time region.

Fits with the double gaussian model show the line variation of the
continuum-intensity sorted data. Contours of the broad versus narrow
component intensities for the three flux levels are shown in Fig. 9. The
narrow component increases as the continuum increases in a roughly
proportional manner whereas the broad component is not a simple
increasing function of the continuum as the {\sl L}$_1$ data show too strong
broad component. From Fig. 9, the ratio of the broad and narrow
components is consistent between the {\sl H}$_1$ and the {\sl M}$_1$ data
indicating that the line profile is similar there. However, the contour
of the {\sl L}$_1$ data is marginally above the region consistent
with a constant narrow/broad component ratio. In these fits, the
continuum photon index is constant within the 90 per cent errors; $\Gamma
= 1.99\pm 0.05$ in the {\sl H}$_1$ and {\sl M}$_1$ spectra and a slightly 
flatter $\Gamma = 1.92\pm 0.06$ in the {\sl L}$_1$ data.

This result shows that the narrow component has a positive correlation
with the continuum, but the broad component may not, at least when the
flux drops, as the {\sl L}$_1$ data show a marginally stronger broad
component than expected from the correlation. The narrow line behaviour
is consistent with the time ordered data sets shown in the previous
section, although any significant correlation between the broad component
and the continuum is not seen. This difference may be due to effects from
the different data sorting methods. It will become clearer in the next
study where we restrict data to the intermediate flux range.

\subsubsection{Intermediate flux-range data}

Making the intensity-sorted spectra, we exclude two time regions of
$1.3\times 10^5$--$1.7\times 10^5$ s (a major flare in {\it i-3}) and
$3.4\times 10^5$--$3.7\times 10^5$ s. The selected count rate ranges,
averaged count rates, integration time of three spectra are summarized in
Table 6. We fit the double gaussian model and show a contour plot between
the broad and narrow line intensities for each spectrum in Fig 10.
Surprisingly the narrow component is consistent with a constant intensity
in contrast to the correlated behaviour with the continuum in time
ordered data sets (section 3.2; see also Fig. 4). A significant line flux
increase is found in the broad component of the high flux data ({\sl
H}$_2$), and the result in Fig. 10 is consistent with the broad
component correlating with the continuum flux. These are opposite
behaviours to the results from data sets sorted in time order.

The selection for the high flux data ({\sl H}$_2$) here mainly
picked up many brief flares, and the result suggests the broad component
of the line follows the continuum rapidly, in $10^4$ s or less. The
increase of the broad component accompanying the brief flares is opposite
to the decrease during the bright flare. This suggests that a different
process is dominating line production in flares of longer and shorter
time scales.

\begin{table}
\begin{center}
\caption{Intensity-sorted data sets excluding the bright flare and the 
deep minimum. Data are sorted according to the count rate in the 0.5--10 keV
S0 light curve. Selected ranges of count rate are given in the second column.
Averaged count rates from each S0 and S1 detector in the 3--10 keV band
over the integration time are indicated.}
\begin{tabular}{cccc}
Data & Range & 3--10 keV & Exposure \\
& \cps & \cps & $10^3$ s \\[5pt]
{\sl H}$_2$ & $> 1.9$ & 0.372/0.300 & 38.4 \\
{\sl M}$_2$ & 1.6--1.9 & 0.304/0.240 & 41.1 \\
{\sl L}$_2$ & $< 1.6$ & 0.251/0.192 & 42.0 \\
\end{tabular}
\end{center}
\end{table} 

\begin{figure}
\centerline{\psfig{figure=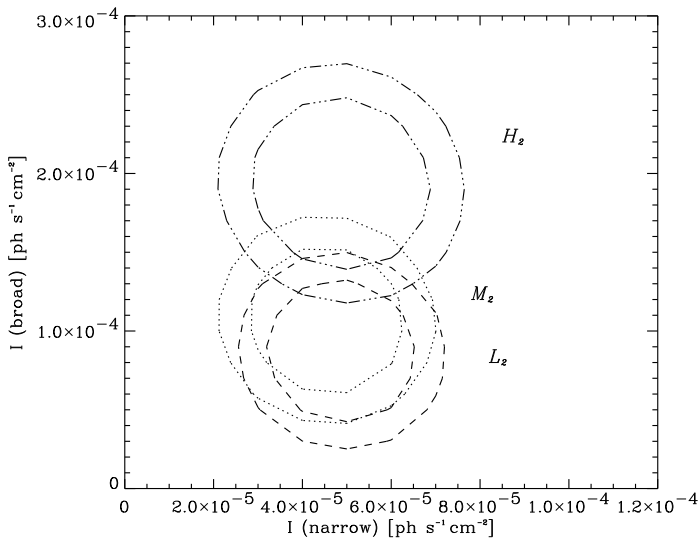,width=0.5\textwidth,angle=0}}
\vspace{0.8cm}
\caption{Contours of intensities in the broad versus narrow components for
intensity-sorted data sets ({\sl H}$_2$; dash-dot line, 
{\sl M}$_2$; dotted line, and
{\sl L}$_2$; dashed line) excluding the time intervals of the 
bright flare and the deep minimum. Contour levels are at 68 and 90 per cent
for the two interesting parameters.}
\end{figure}

\subsection{A count-rate ratio study of rapid variatibility}

Finally, light curves of selected energy bands for the broad red wing
(B; 4.6--6.2 keV), the narrow core (N; 6.2--6.7 keV) and the neighbouring 
continuum (C; 2.4--4.6 keV) are studied with data binned every 4048 s. From 
these light curves, we made ratio plots: 1) the narrow core band
to the continuum band (N/C); 2) the red wing band to the continuum band (B/C);
and 3) the narrow core band to the red wing band (N/B),
averaging the S0 and S1 detectors.

The hypothesis that the data are constant was assesed by the chi-squared
test. Chi-squared values for the best-fit constant model are 77.52 for
N/C = $0.048\pm 0.02$, 182.9 for B/C = $0.321\pm 0.04$, and 71.87
for N/B = $0.146\pm 0.04$, for 84 degrees of freedom, respectively.

Significant variation is found only in the B/C light curve (Fig. 11).
Increases of B/C in the last part of the observation are partly
due to the warm absorber (10--20 per cent; 0.03--0.07 in the B/C).
Nevertheless there is also a decrease during the bright flare. The lack
of any significant result from N/B is mostly due to poor statistics.

\begin{figure}
\centerline{\psfig{figure=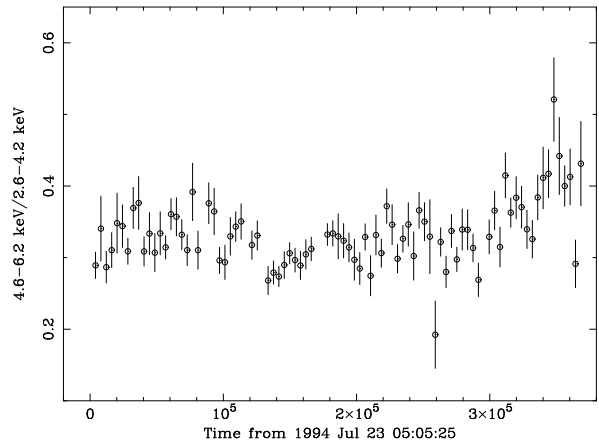,width=0.5\textwidth,angle=270}}
\caption{Time variation of the count-rate ratio of the broad red wing band 
(4.6--6.2 keV) and the neighbouring continuum band (2.6--4.2 keV)
from averaged data from S0 and S1.
Each data point is integrated over 4048 s (some data points have an 
integration time less than 4048 s).}
\end{figure}

\section{DISCUSSION}


\subsection{The black hole and accretion disk in MCG--6-30-15}

MCG--6-30-15 continues to show significant continuum variability (i.e.,
factor of $\sim$ 2 changes) over a wide range of timescale from about 100
s (Reynolds et al 1995) to days. If we assume that the iron line is at
6.4 keV and thus that the disk is not ionized, then the accretion must be
at most a few per cent of the Eddington rate (Ross \& Fabian 1993). The
observed X-ray luminosity (0.5--10 keV) is $\sim 10^{43}$\ergps ~so it is
likely that the mass of the black hole, $10^7M_7$\Ms, is $M_7\geq 1$. For
a typical emission radius of 10$r_1r_g$, with $r_1\approx 1$, the
light-crossing time (of that radius) is then 500$M_7r_1$ s and the
orbital period is $10^4M_7r_1^{3/2}$ s.
Unless $M_7$ is 0.2 or less, the occasional variation on a timescale of
100 s means that the emission region is smaller than the radius, and is
probably a few hot spots above the accretion disk. Relatively small radii
($\leq$ 20 \rg) for the line emitting region implied by the diskline fit
(Tanaka et al. 1995) suggests that the X-ray continuum source should be
just above the disk surface, and orbits around the cental hole with the
disk (Fabian et al. 1995).

\subsection{Variability of the narrow and broad line components}

We detect significant change not only in line flux but also in line
profile (Fig. 4, 5, 6, and 7). A clear result comes from spectral fits to
time-ordered data. Characterizing the iron line as a narrow plus broad
component, the intensity of the narrow component correlates with the
continuum flux whereas the intensity of the broad component possibly
anti-correlates. These line variations are evident in the two time
intervals, the bright flare ({\it i-3}) and the deep minimum ({\it i-7}).
As the EW of the narrow component can be a constant, the broad component
is plausibly responsible for the line shape changes found in this study.
This is consistent with the study of the light curves of the broad
line/continuum bands and the narrow line/continuum bands (section 3.4 and
Fig. 11).

Since the light-crossing time of the line-emitting radius is shorter than
each interval, we do not expect to detect any lag of the line. The
correlated variability of the narrow line component is then reasonable,
although the behaviour of the broad component is difficult to understand.

However, results from the intensity-sorted data suggests that this
behaviour of the line components is no longer true, on taking shorter
time scales. The first attempt using the whole dataset (section 3.3.1) is
largely affected by the extreme lines during the above two time
intervals. Discarding the bright flare and the deep minimum, the
intensity-sorted spectra provides clues to line variability accompanying
continuum change on shorter time scales of less than $10^4$ s (section
3.3.2). The correlation between the narrow component and the continuum
disappears and a clear increase in the broad component during the in high
continuum flux data in which many brief flares are seen.

Summarizing the results of the line variability; 1) the narrow component
followed the continuum if its variation is averaged over a few $10^4$ s,
2) typical continuum changes on short time scale (from $10^3$ s to $10^4$
s) appear to be followed by an immediate response of the broad component
but not of the narrow component, and 3) the broad component changes in a
different manner when the source changes its flux over long timescales (a
few $10^4$ s) such as seen in the bright flare and the deep minimum,
which makes the line profile special during those time intervals.
Implications from this behaviour are discussed below.

\subsection{Evidence for strong gravity around a Kerr hole}

\begin{figure}
\centerline{\psfig{figure=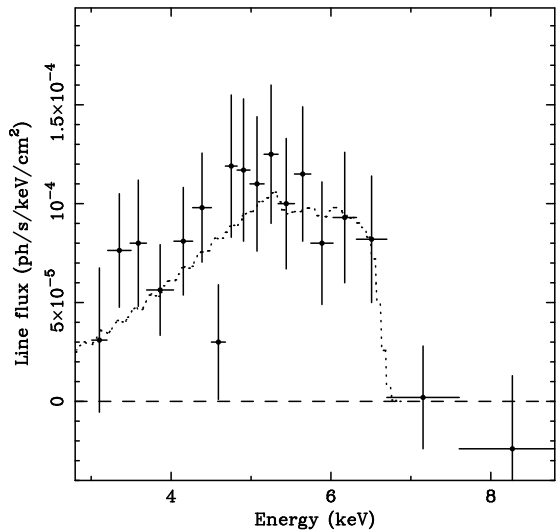,width=0.6\textwidth,angle=270}}
\caption{The line profile obtained from the deep minimum ({\it i-7}).
Data from both S0 and S1 detectors have been summed. The profile has been
corrected for absorption by the 
warm 
absorber and the detector response. The dotted line indicates a
computed diskline model for a Kerr metric (Laor 1991), 
assuming inner and outer radii of $R_{\rm in}\sim1.24$\rg, and $R_{\rm out}
\sim20$\rg of the disk inclined by $i=30$ degree, and a radial emissivity 
index of $\alpha\sim 3$.}
\end{figure}

The huge red tail observed in the deep minimum suggests that the line
should be produced at very small radii (e.g., 1.24--10\rg) requiring the
Kerr geometry of a spinning black hole in which an accretion disk is
dragged into inner radii closer to the black hole than in a Schwarzschild
geometry. Since enourmous gravitational effects are operating at such
small radii, the line from the blue side is also gravitationally shifted
into the red wing. As a result, almost all line emission goes to the
broad red wing (Fig. 12).

The observed EW of this line ($EW\sim 1$ keV) is much larger than
normally expected value even if a reasonable overabundance of iron is
taken into account (e.g., George \& Fabian 1991; Resynolds, Fabian \&
Inoue 1995). Considerable radiation from very small radii of the disk can
however be expected to return to the disk surface by light bending for a
rapidly spinning black hole (returning radiation; Cunningham 1976). It is
therefore possible that reflection will be enhanced by a factor of 2 (see
Table 5 in Cunningham 1976) so that a large equivalent width of the iron
line is produced. Strong reflection is also consistent with the
result of the warm absorber, reflection and the broad line fit (Fig. 8
and Table 4). Another possible reason is an ionized disk. A factor of
$\sim 2$ larger EW for FeXXV than cold iron could occur (Matt, Fabian \&
Ross 1991;
\.Zycki \& Czerny 1992), combined with the returning radiation. A fit
with the diskline model from Laor (1991) for the line energy 6.7 keV
appropriate for FeXXV gives a slightly better fit that the case of 6.4
keV ($\Delta\chi^2 = 1.5$) with a steeper $\alpha = 3.4^{+0.6}_{-0.8}$
(see Table 3).

\subsection{The line profile of the bright flare}

On the other hand, the line shape is dominated by the narrow core around
6.4 keV during the bright flare ({\it i-3}). Here the narrow component
follows the continuum whilst the broad component does not (section
3.2.3). The unusually high narrow-core/red-wing ratio does not fit the
diskline model. A possible explanation for this is given here. The
brightening of the continuum in this interval is of rather a longer time
scale ($\sim 30$ ks) than the other many brief flares in which X-ray flux
increases/decreases on time scales typically less than 10 ks. Suppose
that during this interval most of the activity is from a large flare at
say 7\rg occurs above the blue (approaching) side of the disk. (A ring at
this radius still has a large blue horn.) The line then comes mostly from
the blue side of the disk where the blue peak dominates and the red wing
of the line is depressed.  As noted by Fabian et al. (1995), the effects
of beaming from orbital motion of the disk on the line and continuum
could help slightly here. The beaming factor for the photon rate $\sim
(1+z)^3$ where $z$ is the redshift, is about $\pm30$ per cent at a radius
of 7\rg ~for a disk inclined by 30 degree. The expected slightly larger
EW of the narrow component is compatible with the observed EW (Fig. 4).
In this case, the duration of the bright flare ($\sim 30$ ks) should be
about half an orbit of the accretion disk or less, which implies
$M_7r_1^{3/2}\geq 6$. Given a continuum flaring radius of $\sim 7$\rg
($r_1\sim 1$), the mass of the black hole is then about $6\times 10^7$\Ms
or more. (If the mass is less than this, the result could be due to a
succession of flares on the approaching side of the disk.)

\subsection{Line variability on long and short time scales}

The variability of the both broad and narrow components of the line is
found to be separated into long and short timescales (section 4.2). This
suggests we are seeing different features of line production. Especially,
the behaviour of the broad component in the bright flare and the deep
minimum are remarkably distinct from the others as discussed above. The
anticorrelation between the broad component and continuum seen in the
time-ordered data might be a chance occurrence and requires confirmation
by future observations.

The different line response depending on timescale can be related to
where the continuum and the line are produced. Rapid continuum changes
such as due to many distinct brief flares could occur mainly at small
radii ($r<6r_{\rm g}$) where only a broad component is produced so that
the broad component follows the flares. The constancy of the narrow
component on short timescales may then be due to a constant background of
emission from larger radii (say $6-10 r_{\rm g}$), which changes only on
longer timescales. (If instead the lack of instant response of the narrow
line is due to the time lag effect, the black hole mass can be
constrained as a function of $r_1$ and a typical time scale of the
continuum changes, e.g. if a typical time scale of each brief flare is
about 5000$\tau_5$ s, then the light crossing time of the line emitting
radii 500$M_7r_1$ is larger than that giving $M_7r_1\geq 10\tau_5$.) In
this picture the bright flare is due to a flare or succession of flares
on the blue side of the disk at $\sim 7.5 r_{\rm g}$ and the deep minimum
is due to a simultaneous reduction in the background of outer emission
($r>6r_{\rm g}$) and an increase of inner emission ($r<6r_{\rm g}$).

The net result on line variability is that the line is variable but
complicated and we have found no simpler single explanation other than a
moving pattern of emission from a highly relativistic disk. Complicated
line variability may result from multiple X-ray sources flaring at
different radii on the disk and involve orbital motion of the disk. A
larger detector (or a much brighter source) is required in order to make
firm progress.

\subsection{Comparison with other observations}

We note that a similar relationship between equivalent width of iron-K
line and continuum flux has been found in another broad skewed line
object IRAS~18325--5926 (Iwasawa et al. 1996). The EW is 500--800 eV
during the ASCA observation, higher than the $EW = 390\pm 150$ eV during
the Ginga observation (Iwasawa et al. 1995). The larger value of
equivalent width of the ASCA data was obtained at an averaged flux level
a factor of 3 lower than in the Ginga observation. 
It generally
appears that a large EW tends to be observed when a source is faint
(e.g., Iwasawa \& Taniguchi 1993).

\subsection{Contribution from a torus}

We now comment on reflection from the putative molecular torus. Since the
torus is thought to be outside the broad line region, $\sim$ 1 pc away
from the central source, the torus emission basically remains constant at
a certain averaged intensity level, whereas the reflection from the disk
is supposed to follow the continuum source immediately. If there is a
torus, when the source is faint such as in the deep minimum, then the
torus emission should be pronounced. Such a reflection spectrum is
inevitably accompanied by a sharp 6.4 keV emission line (e.g., George \&
Fabian 1991), as observed in the ASCA spectrum of NGC~2992 (Weaver et al.
1996), and incompatible with the observed spectrum of the deep minimum.
An upper limit of the narrow line equivalent width is $EW\leq 196$ eV, of
which $\sim 120$ eV must be taken by the diskline if the narrow core
keeps a constant EW. Taking account that the average flux level is
greater by a factor of $\sim 2$ than that in the deep minimum, any line
from a torus imposed on the average continuum of MCG--6-30-15 has $EW\leq
40$ eV. This value is much smaller than $EW\simeq 600$ eV observed in
NGC2992 (Weaver et al. 1996), and implies that the reflection from the
torus has little effect on the continuum of the deep minimum, and that
its solid angle is small ($\Omega/2\pi\leq 1/4$), or the column density
is less than \nH$\sim 10^{23}$\psqcm.

\section{CONCLUSIONS}

The iron K line in MCG--6-30-15 shows significant variability both in
intensity and profile when the continuum source changes. Selecting the
two extreme time-intervals corresponding to a bright flare and a deep
minimum, we find a clear difference from the profile at other,
intermediate, flux levels. At these two times the narrow core clearly
correlates with the continuum flux whereas the broad wing possibly
anti-correlates. Particularly, the line profile of the deep minimum shows
a huge red tail for which the Kerr metric is appropriate. These changes
in line profile are understandable in terms of a relativistic accretion
disk around a spinning black hole. Contrarily, on short timescales of
less than $10^4$ s, we find evidence that the broad wing increases as the
continuum flares, whereas this time the narrow core remains unchanged.
This suggests that different regions produce the line at different times.

\section*{ACKNOWLEDGEMENTS}

We thank all the member of the ASCA team who made this long observation
possible. KI thanks the PPARC, JSPS and the British Council for support. 
ACF thanks the Royal Society for support. CSR thanks PPARC for support.
CO is supported by Special Postdoctoral Researchers Program of RIKEN.
WNB thanks the USA National Science Foundation for support.

\end{document}